\documentclass[twocolumn]{aastex62}

\hypersetup{linkcolor=red,citecolor=green,filecolor=cyan,urlcolor=magenta}

\usepackage{braket}
\usepackage{amsmath}
\usepackage{bm}

\newcommand{\lya}{Ly$\alpha\,$}
\newcommand{\lyb}{Ly$\beta\,$}

\newcommand{\mgii}{\ion{Mg}{2}~}

\defcitealias{Eilers2017a}{E17}

\submitjournal{ApJ}

\shorttitle{First Spectroscopic Study of a Young Quasar}
\shortauthors{Eilers et al.}

\begin{document}

\title{First Spectroscopic Study of a Young Quasar}

\author{Anna-Christina Eilers}
\affiliation{Max-Planck-Institute for Astronomy, K\"onigstuhl 17, 69117 Heidelberg, Germany}
\affiliation{International Max Planck Research School for Astronomy \& Cosmic Physics at the University of Heidelberg}
\author{Joseph F. Hennawi}
\affiliation{Physics Department, University of California, Santa Barbara, CA 93106-9530, USA}
\author{Frederick B. Davies}
\affiliation{Physics Department, University of California, Santa Barbara, CA 93106-9530, USA}

\correspondingauthor{Anna-Christina Eilers}
\email{eilers@mpia.de}

\begin{abstract}
The quasar lifetime $t_{\rm Q}$ is one of the most fundamental
quantities for understanding quasar evolution and the growth of
supermassive black holes (SMBHs), but remains uncertain by several
orders of magnitude. In a recent study we uncovered a population of
very young quasars ($t_{\rm Q}\lesssim10^4-10^5$~yr), based on the
sizes of their proximity zones, which are regions of enhanced
Ly$\alpha$ forest transmission near the quasar resulting from its own
ionizing radiation. The presence of such young objects poses
significant challenges to models of SMBH formation, which already
struggle to explain the existence of SMBHs ($\sim
10^{9}\,M_{\odot}$) at such early cosmic epochs.  We conduct the first
comprehensive spectroscopic study of the youngest quasar known, $\rm
SDSS\,J1335+3533$ at $z=5.9012$, whose lifetime is $t_{\rm Q}<10^4$~yr
($95\%$ confidence). A careful search of our deep optical and
near-infrared spectra for \ion{H}{1} and metal absorption lines allows
us to convincingly exclude the possibility that its small proximity
zone results from an associated absorption system rather than a short
lifetime. We use the \mgii emission line to measure the mass of its
black hole to be $M_{\rm BH}=(4.09\pm0.58)\times10^9\,M_{\sun}$,
implying an Eddington ratio of $0.30\pm0.04$ -- comparable to other
co-eval quasars of similar luminosity. We similarly find that the
relationship between its black hole mass and dynamical mass are
consistent with the scaling relations measured from other $z\sim 6$
quasars. The only possible anomaly associated with $\rm
SDSS\,J1335+3533$'s youth are its weak emission lines, but larger
samples are needed to shed light on a potential causality. We discuss the
implications of short lifetimes for various SMBH growth and formation
scenarios, and argue that future observations of young quasars
with JWST could distinguish between them.
\end{abstract}


\section{Introduction}

Quasars are among the most luminous objects in the universe and thus can be observed at very early cosmic epochs, providing unique insights into the initial phases of structure and galaxy formation. Observations indicate that they host super massive black holes (SMBHs), i.e. $M_{\rm BH}\sim 10^9-10^{10} M_{\odot}$ \citep{Mortlock2011, Venemans2013, DeRosa2014, Wu2015}, in their center already at $z\gtrsim 6$, i.e. less than $\sim 1$~Gyr after the Big Bang. The formation and growth of these SMBHs at very early cosmic epochs is a crucial yet unanswered question in studies of black hole and galaxy evolution.

It has been argued that in order to grow the observed masses of SMBHs
this early, very massive initial seeds are required, and, assuming
accretion at the Eddington limit, that accretion must occur
continuously for the entire age of the universe \citep{Volonteri2010, Volonteri2012}. These general considerations imply that the quasar lifetime -- the total integrated time that galaxies shine as quasars --  must be of the order of the Hubble time.

However, measurements of quasar lifetimes have proven to be extremely challenging. 
At low redshifts, i.e. $z\sim 2-4$, quasar lifetimes can be
constrained by comparing the number density of quasars to their host
dark matter halo abundance inferred from clustering studies
\citep{HaimanHui2001, MartiniWeinberg2001, Martini2004,
  WhiteMartiniCohn2008}. But to date this method has yielded only weak constraints on $t_{\rm Q} \sim 10^6-10^9$~yr owing to uncertainties in how quasars populate dark matter halos \citep{Shen2009, White2012, ConroyWhite2013, Cen2015}. Another possibility to infer the lifetime of quasars has been presented by \citet{Soltan1982}, who pointed out that the luminosity function of quasars as a function of redshift reflects the gas accretion history of local remnant black holes. Following this argument \citet{YuTremaine2002} estimate the mean lifetime of luminous quasars from early-type galaxies found in the Sloan Digital Sky Survey (SDSS) to be $t_{\rm Q}\sim 10^7-10^8$~yr. An upper limit on the quasar lifetime, $t_{\rm Q} < 10^9$~yr, is set by the observed evolution of the quasar luminosity function, since the whole quasar population rises and falls over roughly this timescale \citep{Osmer1998}. 
  
Additional uncertainty on the quasar lifetime estimate arises due to the unknown fraction of obscured quasars at high redshift. By analyzing the transverse proximity effect in the \ion{He}{2} Ly$\alpha$ forest, 
 i.e. the transmitted flux of background quasars in the presence of foreground quasars along the line-of-sight, \citet{Schmidt2017} found a bimodal distribution of quasar emission properties, where one population could be relatively old and unobscured, whereas the other could either be younger or highly obscured. 

Several studies of quasars in the local universe suggest that quasars exhibit variability in their activity on timescales ranging from days to $\sim 10^4$~yr, and hence they might accrete matter onto their SMBH in multiple short activity bursts, called the episodic lifetime.
So-called changing-look quasars for instance show variability in their photometric and spectral properties on timescales of days to decades, which has been attributed to changes in the accretion rate of the quasar \citep[e.g.][]{LaMassa2015, Runnoe2016, McElroy2016}, although variable obscuration of the quasar cannot be excluded conclusively.
Evidence for quasar variability on longer timescales of $\sim 10^4$~yr is based on the presence of relic emission nebulae around presumed quiescent galaxies \citep{Schawinski2015, Sartori2016, Keel2017}, or the existence of high-ionization absorption lines in the circumgalactic medium of galaxies in the Cosmic Origins Spectrograph (COS)-halos survey  \citep{Oppenheimer2018}. Such variable quasar activity models, however, cannot explain the ubiquitous presence of large \lya nebulae around quasars at $z\sim 2$ extending out to several hundred kpc (\citealt{Cantalupo2014, Hennawi2015, Cai2018}; Arrigoni-Battaia et al. in prep.), which require a continuous quasar activity for $t_{\rm Q} \sim 10^5-10^6$~yr. 

\begin{figure}[t!]
\centering
\includegraphics[width=.45\textwidth]{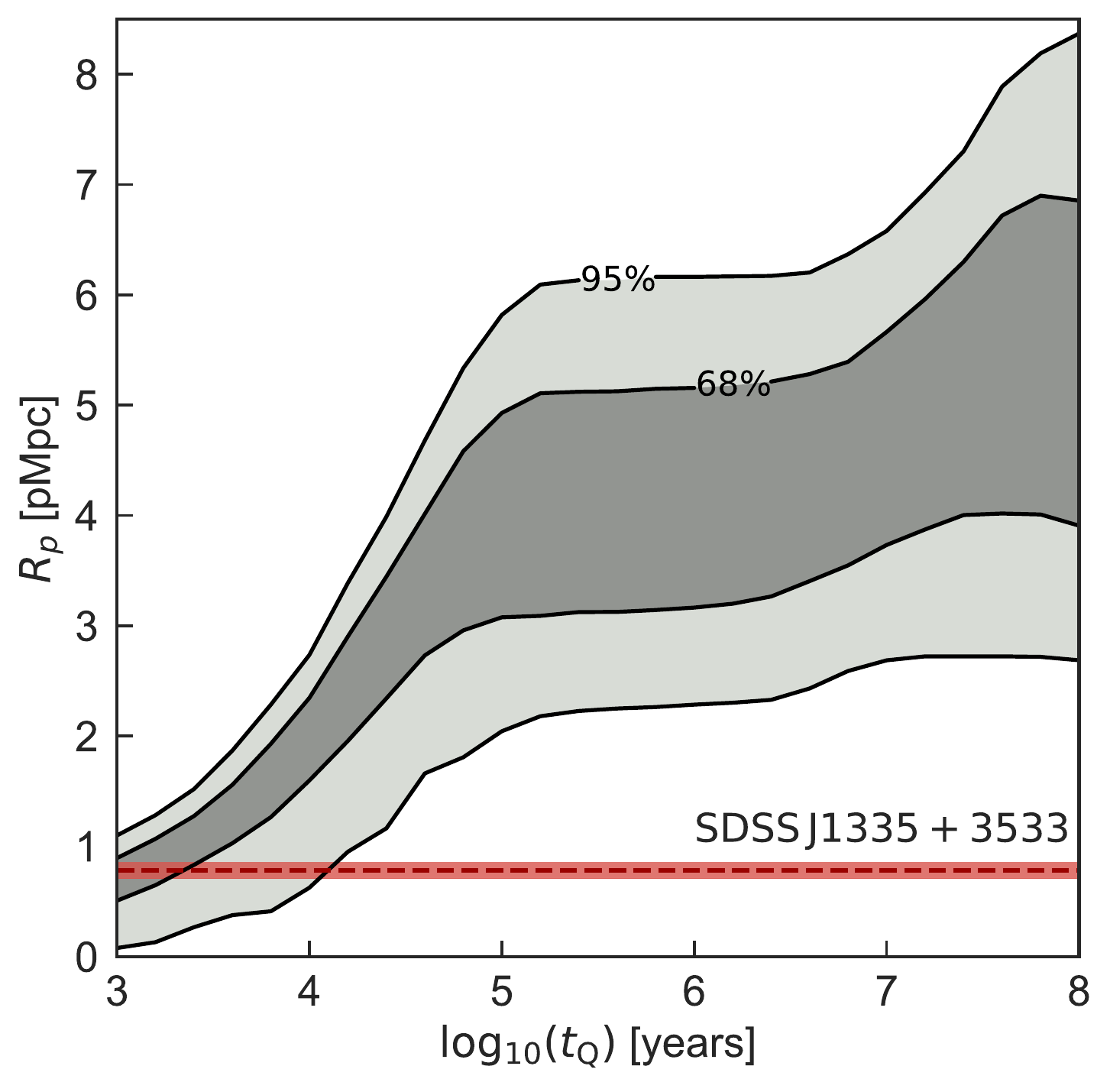}
\caption{Dependence of the proximity zone size $R_p$ on quasar lifetime $t_{\rm Q}$ from radiative transfer simulations for quasars with similar properties as our young object $\rm J1335+3533$, i.e. $z=5.9012$ and magnitude $M_{\rm 1450}=-26.67$: For short lifetimes, i.e. $t_{\rm Q} \lesssim t_{\rm eq}$, the proximity zone grows with lifetime, whereas for longer lifetimes, i.e. $t_{\rm Q} > t_{\rm eq}$, the proximity zone ``saturates'' and ceases to grow. For very long lifetimes of $t_{\rm Q} \gtrsim 10^{7.5}\,$yr, the proximity zones grow slightly again due to heating by the reionization of \ion{He}{2}. The red dashed line indicates the proximity zone measurement of the young quasar $\rm J1335+3533$, i.e. $R_p=0.78\pm0.15$~pMpc, and its uncertainties \citepalias{Eilers2017a}. \label{fig:life}} 
\end{figure}

We recently showed how constraints on quasar lifetimes can be inferred for quasars at high redshift, i.e. $z\gtrsim 6$, which are essential in order to put constraints on theoretical models for the formation and growth of the first SMBHs. The proximity zone, which is the region of enhanced transmitted flux in the Ly$\alpha$ forest in the immediate environment of the quasar owing to its intense ionizing radiation \citep[e.g.][]{Bajtlik1988, MadauRees2000,
  CenHaiman2000, HaimanCen2001, Wyithe2005, BoltonHaehnelt2007a, BoltonHaehnelt2007b, Lidz2007, Bolton2011, Keating2015}, is sensitive to the quasar lifetime, because the intergalactic medium (IGM) has a finite response time to the quasars' radiation, i.e. the gas reaches its new ionization equilibrium state after a timescale $t_{\rm eq} \approx \Gamma^{-1}_{\rm HI}\approx 3\times 10^4$~yr, where $\Gamma_{\rm HI}$ denotes the photoionization rate \citep[][hereafter \citetalias{Eilers2017a}]{Eilers2017a}. 
This dependency of the proximity zone size $R_p$, which is defined as the distance from the quasar to the location where the transmitted flux smoothed to a resolution of $20$~{\AA} in the observed wavelength frame first drops below $10\%$ \citep{Fan2006}, on quasar lifetime inferred from radiative transfer simulations \citep{Davies2016} is shown in Fig.~\ref{fig:life}. 

By analyzing the proximity zones of an ensemble of $34$ high redshift quasars, we recently identified three objects that exhibit extremely small proximity zones, indicating very short quasar lifetimes of $t_{\rm Q}\lesssim 10^5$~yr \citepalias{Eilers2017a}. One of those young quasars $\rm SDSS\, J1335+3533$ (hereafter $\rm J1335+3533$) at $z = 5.9012$, exhibits a particularly small proximity zone, i.e. $R_p = 0.78\pm0.15\,$pMpc, shown as the red dashed line in Fig.~\ref{fig:life}, implying a quasar lifetime of only $t_{\rm Q}\lesssim10^4$~yr. 

In this paper we conduct the first detailed spectroscopic study of the young quasar $\rm J1335+3533$. In \S~\ref{sec:data} we present new optical and near-IR data on the young quasar. We discuss spectral features of the spectrum, measure its black hole mass, and compare its properties to other co-eval quasars of similar luminosity in \S~\ref{sec:spectrum}. 
We show how our new data rule out any other interpretations of its small proximity zone, such as a premature truncation due to close associated absorption systems in \S~\ref{sec:absortion_systems}. 
We summarize our results and discuss several scenarios that could potentially explain the rapid growth of the SMBH in the center of $\rm J1335+3533$ in \S~\ref{sec:summary}, before highlighting a possibility of how we will be able to distinguish between these different scenarios with the upcoming capabilities of JWST. 

\begin{figure*}[t!]
\centering
\includegraphics[width=\textwidth]{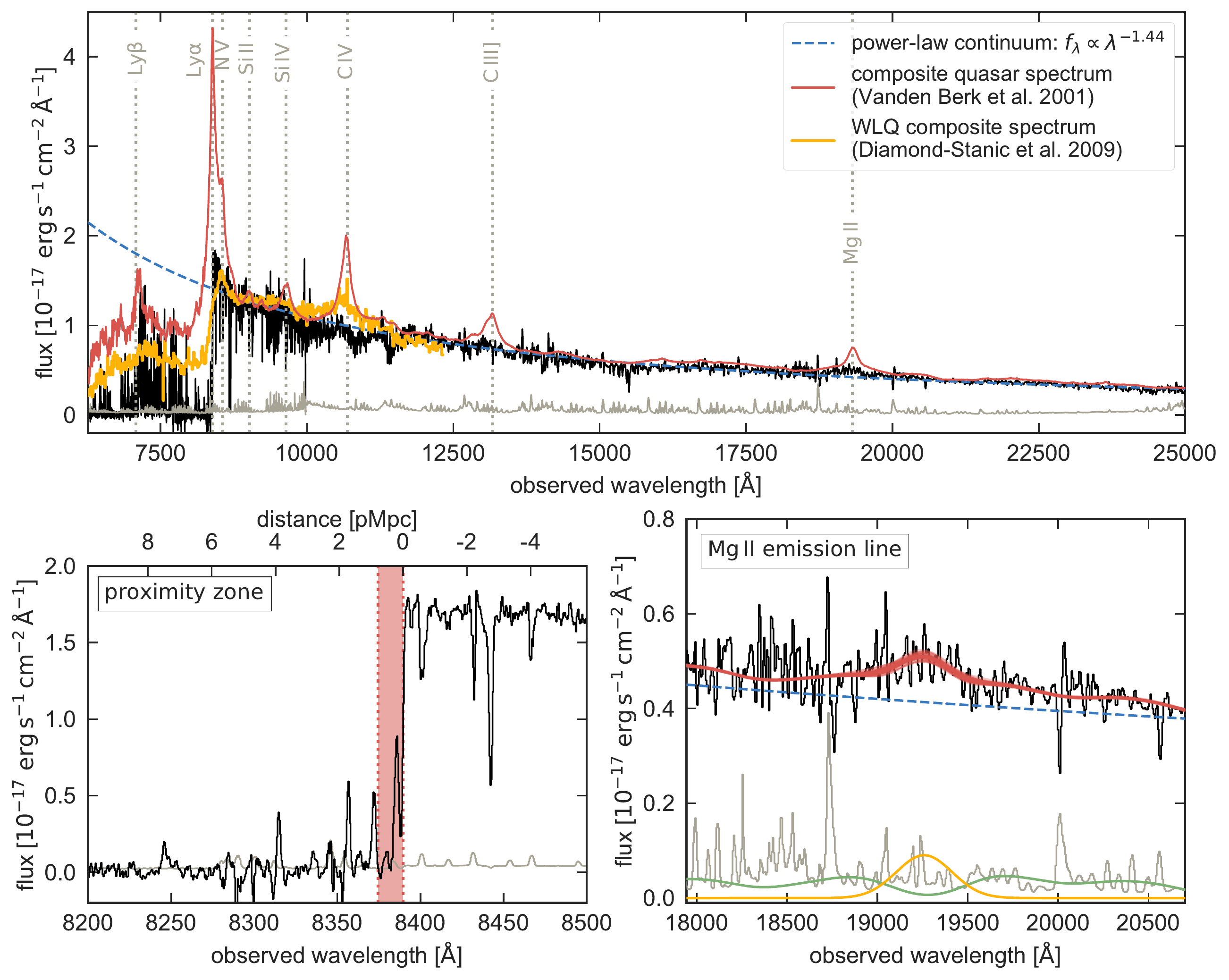}
\caption{\textit{Top panel:} The new optical and near-IR spectrum of $\rm J1335+3533$ and its noise vector are shown in black and grey, respectively. A composite quasar spectrum from low-redshift quasars \citep{VandenBerk2001} is overplotted in red, a WLQ composite spectrum is shown in yellow \citep{DiamondStanic2009}, as well as a power-law continuum estimate of $\rm J1335+3533$, which is shown as the blue dashed line. \textit{Lower left panel:} Zoom-in into the proximity zone of the quasar indicated by the red shaded area. \textit{Lower right panel:} Zoom-in into the spectral region around the \mgii emission line. The spectrum and its noise vector are shown in black and grey, respectively. The quasar continuum fit and its individual components are shown as the colored curves: a power-law continuum from the nuclear emission of the quasar (blue dashed curve), the smoothed iron template spectrum to mimic the emission from the BLR (green curve), a Gaussian function to estimate the width of the broad \mgii emission line (yellow curve), and the total continuum estimate (red curve). \label{fig:spec}} 
\end{figure*}

\section{Observations}\label{sec:data}

In this section we present the new optical (\S~\ref{sec:optical}) and near-IR (\S~\ref{sec:NIR}) data obtained for the young quasar $\rm J1335+3533$. 

\subsection{Optical Spectrum from Keck/DEIMOS}\label{sec:optical}

We acquired three exposures of $20$ minutes each resulting in a total exposure time of $1$ hour on May 26, 2017 (PI: Hennawi), with the Deep Imaging Multi-Object Spectrograph (DEIMOS) at the Nasmyth focus on the Keck II telescope.
We used a custom-made slitmask with a $1\arcsec$ slit and the $830$G grating, resulting in a pixel scale of $\Delta\lambda = 0.47\rm{\AA}$ ($\Delta v \approx 16\rm\, km\,s^{-1}$) and a spectral resolution of $R\approx 2500$. 
The grating was tilted to a central wavelength of $8100\rm{\AA}$ such that the wavelength coverage of the spectrum ranges from $6260\rm{\AA}\leq\lambda_{\rm obs}\leq 10080\rm{\AA}$. 
We reduced the data using standard techniques with the open source python code \texttt{PYPIT}\footnote{\url{http://pypit.readthedocs.io/en/latest/}}, which is available on GitHub\footnote{\url{https://github.com/PYPIT/PYPIT}}. 

\subsection{Near-IR Spectrum from Gemini/GNIRS}\label{sec:NIR}

We obtained the near-IR spectrum of $\rm J1335+3533$ from the Gemini Observatory Archive\footnote{\url{https://archive.gemini.edu}}. The quasar was observed with the GNIRS instrument with the $32\,\rm l/mm$ grating and the short camera's cross-dispersion prism on January 27, 2016, in a program aiming to obtain near-IR data for $60$ quasars at $z>5.7$ to study their physical properties, such as black hole masses, metallicities and absorption features (Program: GN-2016A-LP-7, PI: Shen). 
The data was collected following an ABBA dither pattern using twelve $300\rm\,s$ individual exposures, resulting in a total of $1$ hour exposure time. The spectrum covers the wavelength range between $0.85-2.5\,\mu$m with a resolution of $R\approx 1800$. 
The data has been reduced making use of the Low-Redux pipeline\footnote{\url{http://www.ucolick.org/~xavier/LowRedux/}} developed as part of the XIDL\footnote{\url{http://www.ucolick.org/~xavier/IDL/}} suite of astronomical routines in the Interactive Data Language (IDL), which employs standard data reduction techniques. 

The \mgii emission line of the quasar, which we use to measure its black hole mass (see \S~\ref{sec:mass}), lies in a spectral region with substantial telluric absorption. We followed the telluric standard star correction procedure that is part of the XIDL pipeline in order to minimize the effects of atmospheric absorption. This procedure compares the telluric standard star taken before and after the observation to models of stellar spectra, thus deriving both the telluric correction and the flux calibration simultaneously. We apply the telluric corrections using both standard stars and found that both standard stars yield the same results. As an additional check, we also applied one telluric standard correction to the other standard star and found that our pipeline was producing good corrections in the wavelength range around \mgii. 

Our new optical and near-IR data of $\rm J1335+3533$ is shown in the top panel of Fig.~\ref{fig:spec}, as well as a zoom-in onto the region around the extremely small proximity zone (lower left panel). We normalized the GNIRS data to match the $J-$band magnitude $J_{\rm AB} = 19.84$ and re-scaled the DEIMOS spectrum slightly
to match the near-IR data at $\lambda\approx 10,000\rm{\AA}$.

\section{Spectral Features of the Young Quasar $\rm J1335+3533$}\label{sec:spectrum}

The new data presented in the previous section shows that $\rm J1335+3533$ exhibits very weak emission lines (\S~\ref{sec:wel}). 
The near-IR spectrum covering the \mgii emission line enables us to estimate the mass of the central SMBH (\S~\ref{sec:mass}) of $\rm J1335+3533$ and to compare the quasar's properties to other quasars at similar redshift (\S~\ref{sec:scaling}).

\subsection{Weak Emission Lines}\label{sec:wel}

A comparison between the spectrum of $\rm J1335+3533$ and a composite quasar spectrum created from $>2200$ quasar spectra at redshifts $0.044\leq z \leq 4.789$ from SDSS \citep{VandenBerk2001} as well as composite spectrum of $32$ weak emission line quasars (WLQs) at $z>3$ \citep{DiamondStanic2009} shown in the top panel of Fig.~\ref{fig:spec}, clearly reveals the lack of strong emission lines in the spectrum of $\rm J1335+3533$, 
as previously noticed by \citet{Fan2006_discovery}. 
We measure an equivalent width (EW) of EW(Ly$\alpha\,+\,$\ion{N}{5})$\approx5.1\rm\,{\AA}$ by fitting two Gaussian functions to the red side of the Ly$\alpha\,+\,$\ion{N}{5} emission lines at $\lambda_{\rm rest}\geq 1215.67\rm{\AA}$, which falls into the category of a WLQ according to the definition by \citet{DiamondStanic2009}, who defined WLQs as quasars with EW(Ly$\alpha\,+\,$\ion{N}{5})$<15.4\,\rm{\AA}$. 

The top panel of Fig.~\ref{fig:spec} shows a power-law continuum fit to all wavelengths $\lambda_{\rm rest} \geq 1215.67\rm{\AA}$, with an estimated slope $\alpha_\lambda = -1.444\pm 0.002$, which is consistent with other quasar spectra \citep{VandenBerk2001, Lusso2015, DiamondStanic2009}.

\subsection{Measurement of the Black Hole Mass}\label{sec:mass}

We derive the black hole mass from the single-epoch near-IR spectrum covering the \mgii$\lambda2798.7\,${\AA} emission line assuming that the dynamics in the quasar's broad line region (BLR) are dominated by the gravitational pull of the black hole and the virial theorem can be applied. To estimate the black hole mass we use the relation
\begin{equation}
\frac{M_{\rm BH}}{M_{\sun}} = 10^{6.86}\left(\frac{\text{FWHM}_{\rm Mg\,II}}{10^3\,\rm km\,s^{-1}}\right)^2\left(\frac{\lambda L_{\lambda,\,3000\text{\AA}}}{10^{44}\,\rm erg\,s^{-1}}\right)^{0.5}\label{eq:mbh}
\end{equation}
which has been calibrated by \citet{VestergaardOsmer2009} to pre-existing black hole mass scaling relations using the FWHM of other emission lines, such as H$\beta$ or \ion{C}{4}, by means of several thousand high-quality spectra from the Sloan Digital Sky Survey (SDSS). 

In order to obtain an estimate of the black hole mass of $\rm J1335+3533$, we model the quasar emission within the wavelength region around the \mgii line of the quasar, i.e. $2100\,\rm{\AA}\,\leq \lambda_{\rm rest}\leq 3088\,\rm{\AA}$, as a superposition of a power-law continuum arising from the quasar's nucleus, a scaled template spectrum of the emission from the iron lines, \ion{Fe}{2} and \ion{Fe}{3}, arising from the BLR of the quasar and a single Gaussian function to model the \mgii emission line, i.e. 
\begin{equation*}
f_{\lambda} = f_0 \lambda ^{-\alpha_{\lambda}} + f_1\cdot f_{\lambda,\,\rm iron} + f_2 \cdot \exp\left(-\frac{(\lambda - \mu_{\rm MgII})^2}{2\sigma_{\rm MgII}^2}\right)
\end{equation*}
We make use of the iron template spectrum from \citet{VestergaardWilkes2001} and convolve the original template derived from a narrow emission line quasar with a Gaussian kernel to adapt it to the width of the broad emission lines in the spectrum of $\rm J1335+3533$\footnote{The width of the Gaussian smoothing kernel $\sigma_{\rm conv}\approx 2089\,\rm km\,s^{-1}$ is determined by eqn.~$1$ in \citet{VestergaardWilkes2001}, assuming a $\rm FWHM \sim 5000\,\rm km\,s^{-1}$ of the quasar's emission lines. }.

We apply the MCMC affine-invariant ensemble sampler \texttt{emcee}\footnote{\url{http://dfm.io/emcee/current/}} \citep{emcee} 
to fit the quasar spectrum, where our likelihood is $-2\ln \mathcal{L}= \chi^2$ making use of the noise vector of the spectrum. We assume flat priors for the free parameters, i.e. $f_0\in[0, 10]$, $\alpha_{\lambda}\in[-2, 0]$, $f_1\in[0, 1]$, $f_2\in[0, 1]$, $\sigma_{\rm MgII}\in[0, 1000]$, $\mu_{\rm MgII}\in[-1000, 1000]$. We adopt the median of the resulting posterior probability distributions as our parameter estimates. For the slope of the power-law continuum within this small wavelength region around \mgii we obtain $\alpha_{\lambda}=\-1.21\pm 0.02$, which differs slightly from the slope we inferred for the whole spectrum. 
From the power-law continuum we can derive the luminosity $\lambda L_{\lambda,\,3000\text{\AA}} = (3.15\pm 0.05)\times 10^{46}\,\rm erg\,s^{-1}$.

For the Gaussian fit to the \mgii line we obtain a width of $\rm FWHM = 5645 \pm 396\,\rm km\,s^{-1}$ and we measure a significant blueshift of $\Delta v_{\rm MgII} = -815\pm 139\,\rm km\,s^{-1}$ compared to the systemic redshift of the quasar at $z=5.9012$, derived from measurements of the CO (6-5) $3~\rm mm$ line by \citet{Wang2010}. Such large blueshifts of the \mgii line have been observed in other high redshift quasars \citep{Venemans2016}. 
The resulting continuum fit around the \mgii emission line and its individual components are shown in the lower right panel of Fig.~\ref{fig:spec}. 
Given eqn.~\ref{eq:mbh} we estimate the mass of the central SMBH of $\rm J1335+3533$ to be
\begin{equation*}
M_{\rm BH} = (4.09 \pm 0.58)\times 10^9 \,M_{\sun}. 
\end{equation*}
Note that the uncertainty in this measurement does not include the intrinsic scatter of $0.55$~dex in the scaling relation itself \citep{VestergaardOsmer2009}. 

\begin{figure*}[t!]
\centering
\includegraphics[width=\textwidth]{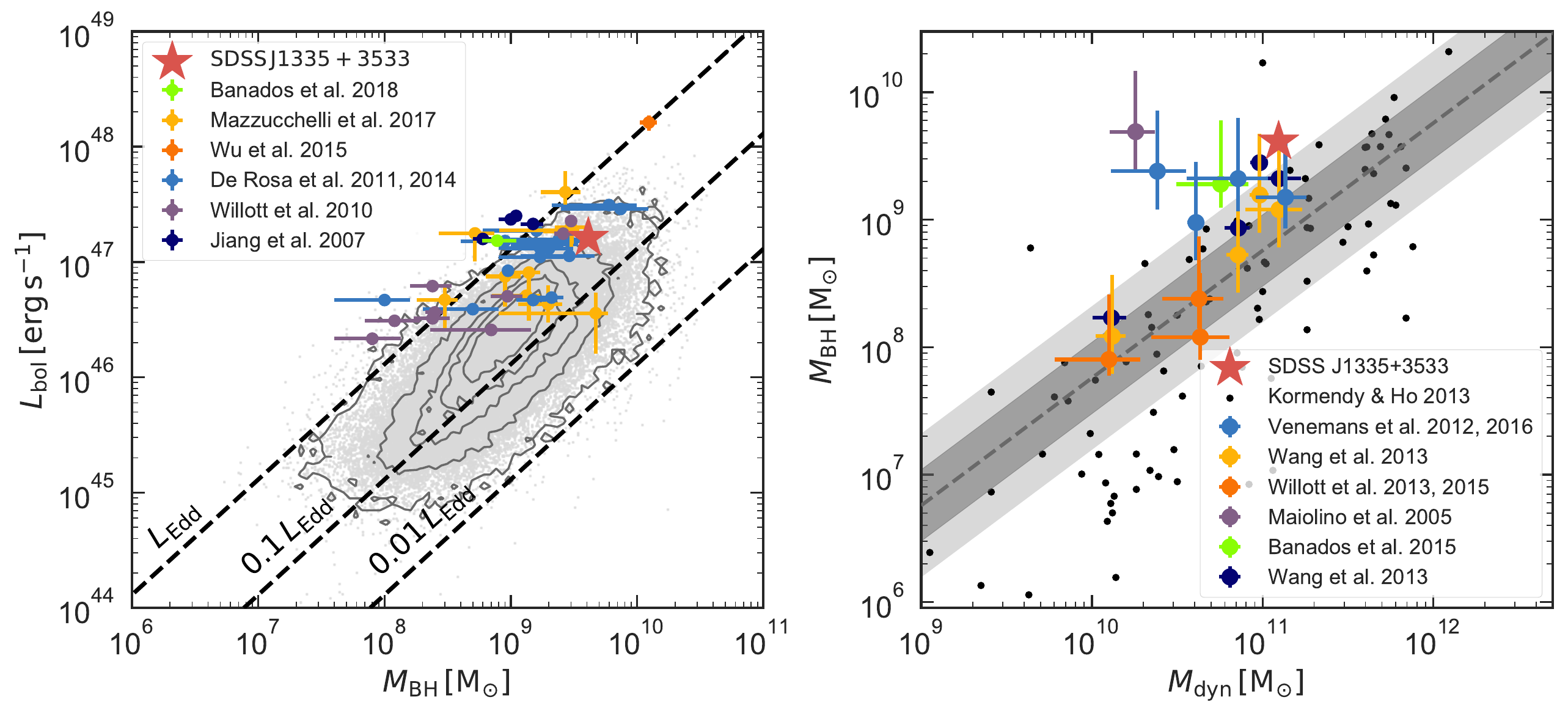}
\caption{\textit{Left panel:} Relation between bolometric luminosity $L_{\rm bol}$ and black hole mass $M_{\rm BH}$ for a sample of low redshift quasars at $0.4\lesssim z\lesssim 2.2$ from the SDSS Data Release 7 \citep{Shen2011} indicated by the grey contours and high redshift quasars, which are shown by the colored data points. The young quasar $\rm J1335+3533$ is indicated as the red star. Note that the error bars here are smaller than the symbol itself. \textit{Right panel:} Relation between the quasar's black hole mass $M_{\rm BH}$ and its dynamical mass $M_{\rm dyn}$. The grey dashed line shows the relation derived from local galaxies by \citet{KormendyHo2013} with its $1\sigma$ and $2\sigma$ uncertainties. $\rm J1335+3533$ is indicated by the red star, whereas other measurements for high redshift quasars are shown as the colored data points. \label{fig:BH}} 
\end{figure*}

\subsection{Black Hole Mass Scaling Relations}\label{sec:scaling}

It is interesting to know how the properties of the young quasar $\rm J1335+3533$ compare to other co-eval quasars with similar luminosities. 
We infer the bolometric luminosity for $\rm J1335+3533$ following \citet{Shen2008} via $L_{\rm bol} = 5.15\,\lambda\,L_{\lambda,\,3000\text{\AA}} = (1.62\pm 0.03)\times 10^{47}\,\rm erg\,s^{-1}$. Given $M_{\rm BH}$ which we estimated in the previous
section, we can compute its Eddington luminosity $L_{\rm Edd} = (5.32\pm 0.75)\times 10^{47}\,\rm erg\,s^{-1}$, indicating that the quasar is accreting with an Eddington ratio of $L_{\rm bol}/L_{\rm Edd} = 0.30 \pm 0.04$. 

In the left panel of of Fig.~\ref{fig:BH} we show the relation between the bolometric luminosity and the black hole mass of $\rm J1335+3533$ in comparison to a low redshift quasar sample of $\gtrsim 75,000$ objects at $0.4\lesssim z\lesssim 2.2$ from the SDSS Data Release 7 \citep{Shen2011, Wang2015}, and in comparison to several other quasars at high redshift, i.e. $z\gtrsim 5.8$ \citep{Jiang2007, Willott2010b, DeRosa2011, DeRosa2014, Wu2015, Mazzucchelli2017, Banados2018}. 
The black dashed curves indicate regions with constant Eddington luminosity \citep{Eddington1922, MargonOstriker1973}. 

We present the scaling relation between black hole mass and dynamical mass of the quasar's host galaxy $M_{\rm dyn}$ in the right panel of Fig.~\ref{fig:BH}. The grey dashed line shows the relation derived by \citet{KormendyHo2013} on which local galaxies (shown as black data points) fall with an intrinsic scatter of $0.28$~dex. 

The dynamical mass of $\rm J1335+3533$ has been derived by \citet{Wang2010} from the FWHM of the CO$(6-5)$ line to be $M_{\rm dyn} \sin^{2}i \approx 3.1 \times 10^{10}\, M_{\sun}$.
 The large $5\arcsec$ beam of their observations ($\approx 29.2$~kpc at $z\sim 6$) does not spatially resolve the source and hence they assumed a quasar host size radius of $R\approx 2.5\,$kpc. 
Assuming an inclination angle of $i=30^{\circ}$, we estimate $M_{\rm dyn} \approx 1.24\times 10^{11}\,M_{\sun}$. 
The colored data points in the right panel of Fig.~\ref{fig:BH} show measurements of other high redshift quasars \citep{Maiolino2005, Wang2013, Willott2013, Willott2015, Venemans2012, Venemans2016, Banados2015}, which show consistently higher black hole masses than the galaxies in the local universe. Note however, that these measurements are highly uncertain since in most cases the quasar's host galaxy is not resolved and its inclination angle is unknown.

In both the Eddington ratio and the $M_{\rm BH}-M_{\rm dyn}$ relation shown in Fig.~\ref{fig:BH} our young object $\rm J1335+3533$, indicated by the red star, does not hold any special position in the parameter space; its black hole mass, bolometric luminosity and dynamical mass are consistent with other quasars at similar redshifts.

\section{Search for Associated Absorption Systems}\label{sec:absortion_systems}

In \citetalias{Eilers2017a}, we argued that the small proximity zone of $\rm J1335+3533$ was unlikely to arise due to a truncation from self-shielding associated absorption systems that are optically thick at the Lyman limit, such as Damped \lya systems (DLAs) or Lyman Limit systems (LLSs) which would block the quasar's ionizing radiation. However, this was based on a spectrum of much lower $\rm S/N$ than our current data. Hence, we revisit this question using our new data, which enables a thorough search for strong associated absorption systems along the line-of-sight to the quasar.

In order to prematurely truncate the proximity zone a self-shielding absorption system would need to be located within $\lesssim 1000\rm\,km\,s^{-1}$ of the quasar, around the edge of the proximity zone, which ends at $z\approx 5.889$. 
To this end, we search for signatures of strong absorption line systems in the quasar spectrum near the end of its proximity zone by visually inspecting the spectrum, i.e. we search for evidence of damping wings which would indicate the presence of a strong DLA absorber, as well as associated ionic metal-line transitions, but do not find either. We also consider where an optically thick absorber, i.e. $\log N_{\rm HI} \geq 17\,\rm cm^{-2}$, that does not exhibit any damping wings because it might have a small column density and undetected metal absorption lines possibly due to very pristine gas \citep{Fumagalli2011, Simcoe2012, Cooper2015, Cooke2017}, could be located by inspecting its location within the Lyman series forests, since the spectrum at the location of an optically thick absorber should be line black everywhere. 

\begin{figure*}[t!]
\centering
\includegraphics[width=\textwidth]{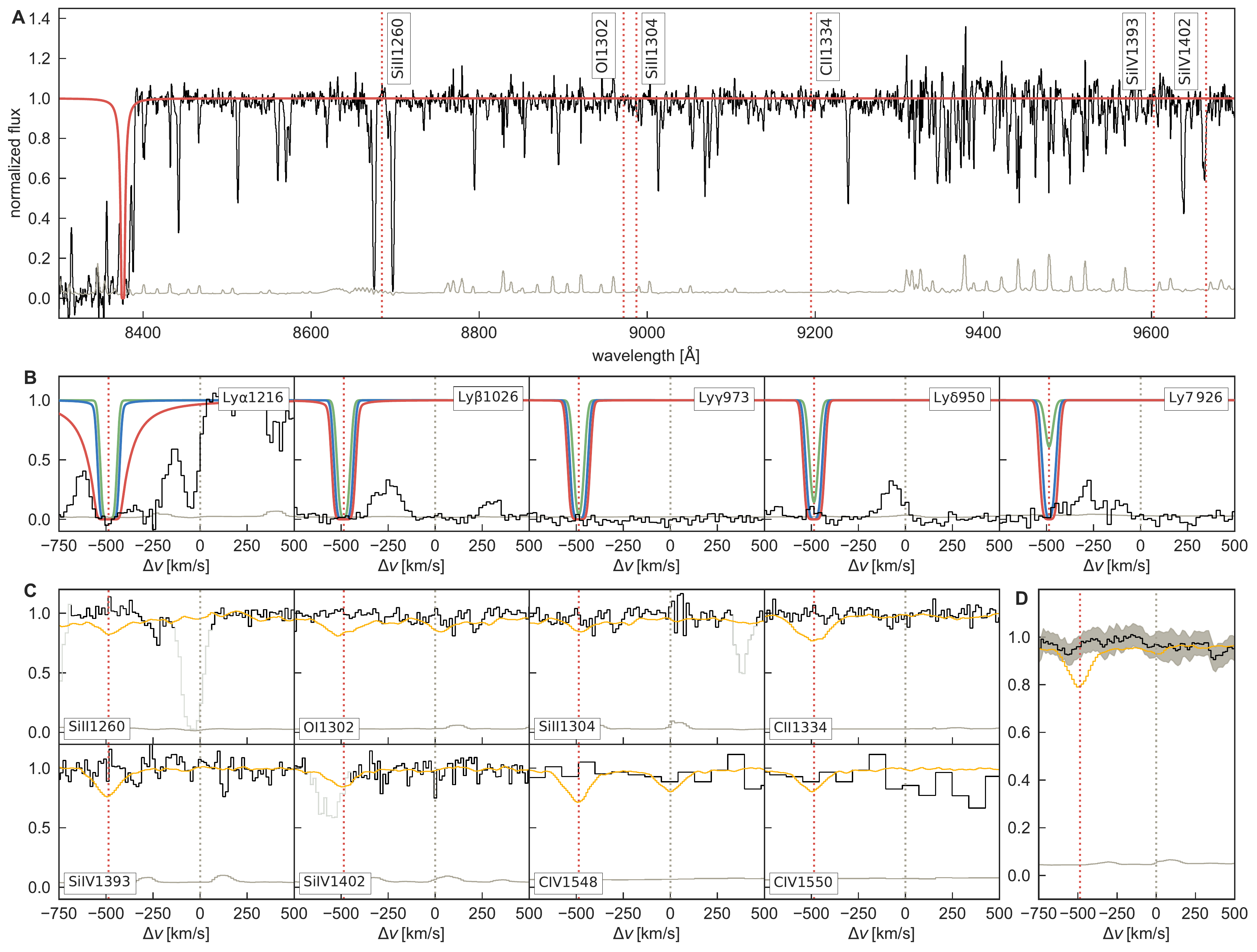}
\caption{\textit{Panel A:} Continuum normalized DEIMOS spectrum of $\rm J1335+3533$ and its noise vector are shown in black and grey, respectively. A possible absorption system ($\log N_{\rm HI} = 18.5\,$cm$^{-2}$, $b=20\,$km$\,$s$^{-1}$) located in front of the quasar ($z_{\rm abs} = 5.89$) is shown as the red curve. The expected positions of metal lines that would be associated with the absorption systems are indicated by the red dashed lines. Note that the two strong absorption lines at $8670\,${\AA} and $8700\,${\AA} are the doublet lines \ion{Mg}{2} $\lambda 2796\,${\AA} and \ion{Mg}{2} $\lambda 2803\,${\AA} that belong to an absorption system at $z_{\rm abs}\approx2.102$. \textit{Panels B:} Three absorption systems with varying column densities ($\log N_{\rm HI} = 18.5\,$cm$^{-2}$ (red), $\log N_{\rm HI} = 17.0\,$cm$^{-2}$ (blue), and $\log N_{\rm HI} = 15.5\,$cm$^{-2}$ (green), $b=20\,$km$\,$s$^{-1}$) smoothed to the spectral resolution of the DEIMOS data within the Lyman series forest. The transmission spikes observed in the $\rm Ly\delta$ and $\rm Ly7$ forest constrain the column density of the potential absorption system. The grey dotted lines mark the position of the quasar. \textit{Panels C:} Zoom-ins into the spectral regions around the expected metal absorption lines. In yellow we show a composite spectrum of $20$ LLSs by \citet{Fumagalli2013} centered around the redshift of the absorber (red dashed lines). The gray colored parts of the spectrum mark absorption features from foreground absorption systems. \textit{Panel D:} Stack of the spectral regions around the expected absorption lines. The grey shaded region shows the $1\sigma$ uncertainty. The yellow curve shows the stack of the composite LLS spectrum. \label{fig:abs}} 
\end{figure*}

In Fig.~\ref{fig:abs} we show a hypothetical absorption system at $z_{\rm abs}=5.89$ that would be able to truncate the proximity zone and indicate the locations where metal absorption lines associated with the absorber would fall in the quasar spectrum (panel A). The redshift of the hypothetical absorption system was chosen at the only position possible in the spectrum, such that it lies just at the end of the proximity zone and does not interfere with the clear transmission spikes observed in the \lya and \lyb forest.  
We show the individual regions around the expected metal absorption lines (panels C) and overplot a composite spectrum of $20$ LLSs with $\log N_{\rm HI} < 19\rm\,cm^{-2}$ by \citet{Fumagalli2013}, but do not find any evidence for the presence of such metal absorption lines in the spectrum that could be associated with a close absorption system, neither in the individual spectral regions 
nor when stacking the spectral regions around the absorption system (panel D). 

Additionally, we examine the location of the absorption system within the Lyman series forests (panels B) and find transmission spikes in the $\rm Ly\delta$ and $\rm Ly7$ forests which set a strong limit on the column density of any potential absorption system. In order to not violate the forest transmission, we find that the column density of an absorption system would have to be $N_{\rm HI}\lesssim 15.5\rm\,cm^{-2}$ (green curves), which would have an optical depth to ionizing photons of $\tau_{\rm HI}\ll1$, which is below the threshold of an optically thick absorber, such as a LLS.
Hence we conclude that the proximity zone of $\rm J1335+3533$ has not been prematurely truncated by an optically thick absorber.

\section{Summary and Discussion}\label{sec:summary}

\begin{figure*}[t!]
\centering
\hspace{.47 cm}
\includegraphics[width=.775\textwidth]{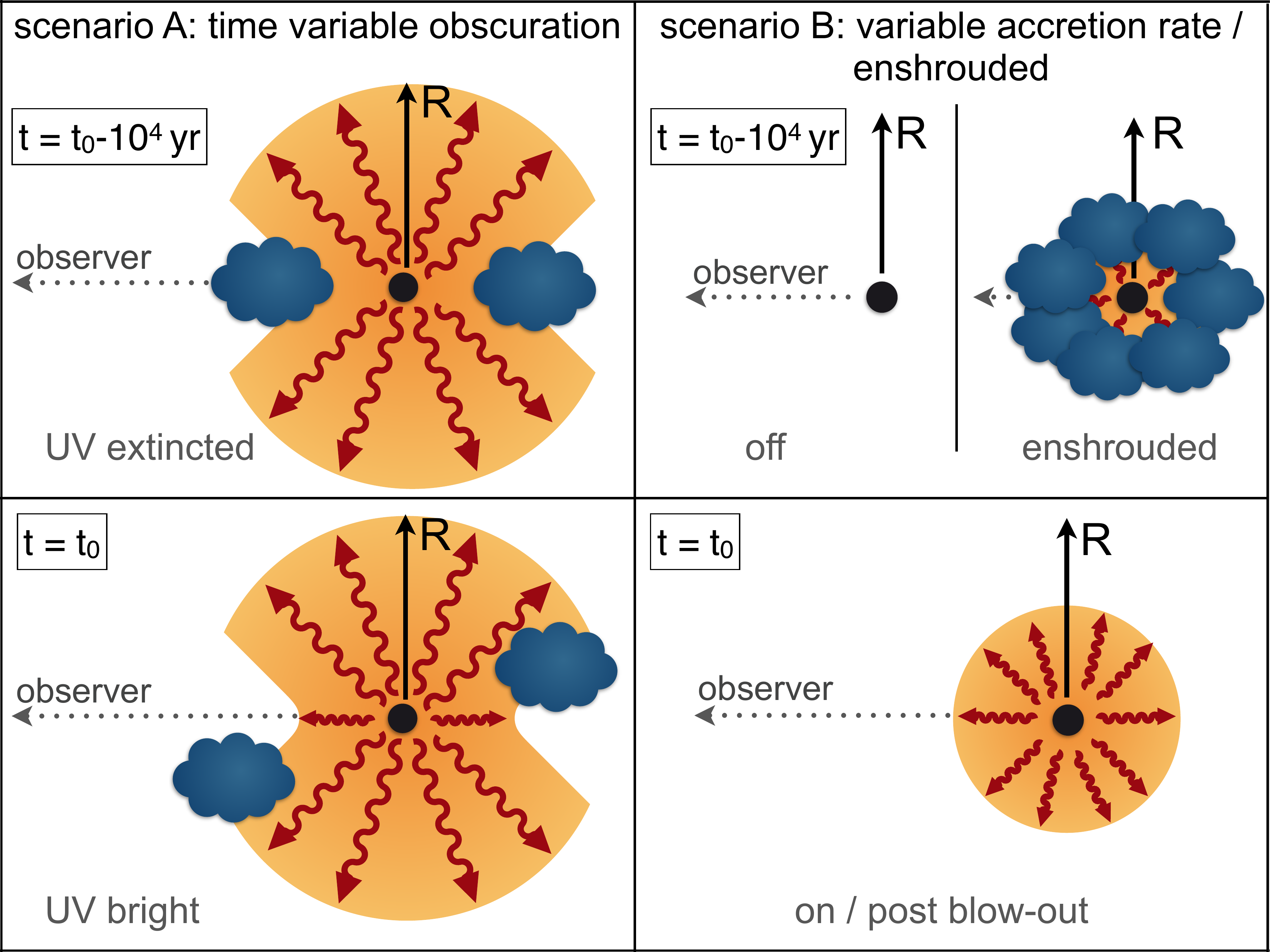}
\includegraphics[width=\textwidth]{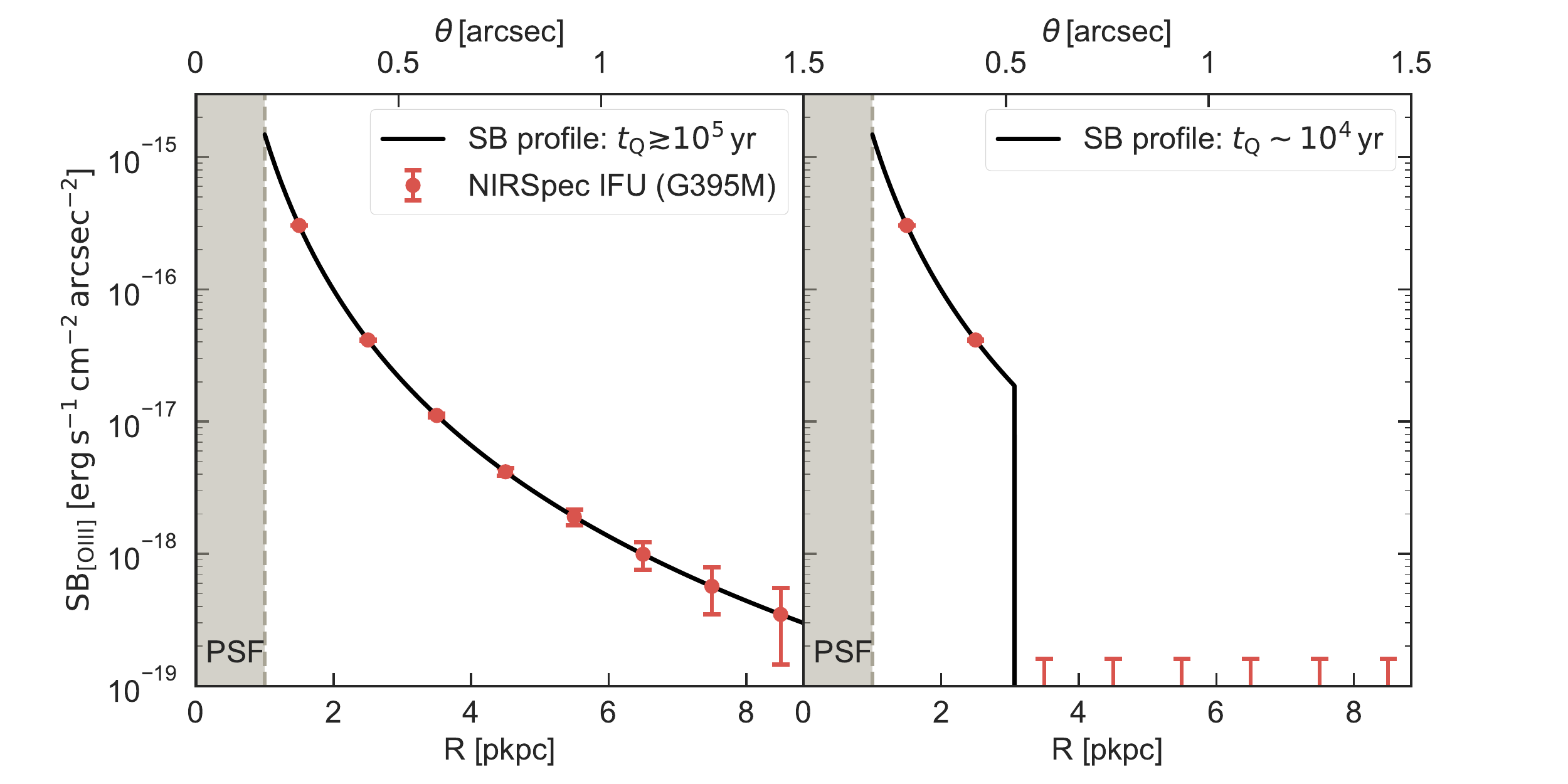}
\caption{\textit{Top panels:} Different black hole growth scenarios that could explain our observations of the young quasar. The left panels show a quasar with a long lifetime, i.e. $t_{\rm Q}\gtrsim 10^5$~yr, that was previously (at a time $t\leq t_0-10^4~$yr) UV obscured along our line-of-sight and is now, at $t=t_0$, UV bright. The right panels show a quasar, whose UV luminous phase just turned on $\sim 10^4~$yr ago, possibly after an initial enshrouded phase. \textit{Bottom panels:} Surface brightness profiles of the ionized ENLR traced by [\ion{O}{3}] we expect to observe around the quasar for the two different scenarios. 
The red data points show the expected signal within a $\sim 3$ hour exposure with NIRSpec IFU on JWST, when averaged over all spatial pixels in annuli of $\Delta R = 1$~kpc around the quasar and assuming $\sim 3$ spectral pixels within the [\ion{O}{3}]$\lambda 5007${\AA} emission line when observed with the medium resolution grating $\rm G395M$.\label{fig:jwst}} 
\end{figure*}

We presented new optical and near-IR spectra of the quasar $\rm J1335+3533$, which we previously identified to be a very young object, i.e. $t_{\rm Q}\lesssim 10^4$~yr, by analyzing the extent of its proximity zone \citepalias{Eilers2017a}. 
With our new high $\rm S/N$ data we can rule out any premature truncation of the proximity zone due to a self-shielding absorption system and constrain the column density of a potential absorber to be $\log N_{\rm HI}\lesssim 15.5\rm cm^{-2}$ due to the presence of flux transmission spikes in the higher order Lyman series forest. 

By measuring the FWHM of the single-epoch \mgii emission line of the quasar we estimate the mass of the central SMBH to be $M_{\rm BH} = (4.09 \pm 0.58)\times 10^9 M_{\sun}$, which is in good agreement with other SMBH mass measurements of co-eval quasars. Other properties of the young quasar and its host galaxy such as the bolometric luminosity, the dynamical mass, the Eddington ratio, and the slope of its continuum emission are likewise in good agreement with other high redshift quasars. 

Broadly speaking, $\rm J1335+3533$ is fully consistent with a typical $z\sim 6$ quasar and shows no evidence for anomalies associated with its young age, with the potential exception of the fact that it is a WLQ. 
\citet{Shemmer2009} discussed whether WLQs could be extreme
objects with very high accretion rates, which could potentially
explain a younger age. Since X-rays probe the innermost regions of
quasars, they studied a sample of $11$ WLQs spanning the redshift range $2.7 < z <
5.9$ with \textit{Chandra} to provide diagnostics of the central
accretion process. 
However, they did not find any evidence for a hard X-ray spectrum that would be a characteristic of high accretion rates. Interestingly, $\rm J1335+3533$ is a member of their sample, for which they found upper limits of $<3.0$ and $<6.4$ counts in the ultrasoft ($0.3-0.5$~keV) and hard X-ray band ($2-8$~keV), respectively, and  $3.0^{+2.9}_{-1.6}$ counts in the soft X-ray band ($0.5-2$~keV) in a $23.47$~ks exposure, suggesting only very modest accretion rates. 

\citet{Hryniewicz2010} also analyze a WLQ at lower redshift 
and argue that despite its fairly low Eddington ratio and line properties that are not consistent with the trends expected for quasars with high accretion rates, the most plausible explanation for its weak emission lines is that the quasar activity has just started and the regions from where broad lines are emitted, did not yet respond to this activity. 

Following the definition for WLQs by \citet{DiamondStanic2009}, i.e. EW(Ly$\alpha\,+\,$\ion{N}{5})$<15.4\,\rm{\AA}$, \citet{Banados2016} estimated that $13.7\%$ of the $117$ Pan-STARRS 1 (PS1) quasars at $z>5.7$ that they analyzed fall in this category, which is higher than the fraction of young quasars we estimated. 
However, out of the three young objects identified in \citepalias{Eilers2017a} not only $\rm J1335+3533$ but also another object, $SDSS\,J0100+2802$, exhibits weak emission lines according to \citep{Wu2015}, although \ion{Si}{4}and \ion{C}{4} emission lines are present, suggesting a potentially higher fraction of WLQs among the population of young quasars than in the average quasar population. 

The measurement of a billion solar mass black hole in the center of $\rm J1335+3533$ poses the question of how the SMBH could have grown to be so massive within a time frame of only $t_{\rm Q}\lesssim 10^4$~yr, when usually at these redshifts quasar lifetimes comparable to the Hubble time with Eddington limited accretion rates have to be invoked in order to explain the existence of SMBHs. 
Assuming a simple light-bulb light curve model, which implies that the quasar shines continuously at the same luminosity, and a fiducial quasar lifetime of $t_{\rm Q}\sim 10^8$~yr, the probability of finding a quasar that turned on just $\sim 10^4$ years ago would be very small, i.e. $\sim 0.01\%$. This discrepancy can be resolved if one assumes a shorter fiducial quasar lifetime of $t_{\rm Q}\sim 10^6$~yr, resulting in a probability of $\sim1\%$ for finding these objects with short lifetimes, which is roughly consistent with the detection rate in our ensemble (in \citetalias{Eilers2017a} we find one quasar, $\rm J1335+3533$, with $t_{\rm Q}\lesssim 10^4~$yr, i.e. $1/34\approx 3\%$, and two additional objects with $t_{\rm Q}\lesssim 10^5~$yr, i.e. $3/34\approx 9\%$). However, an average quasar lifetime of $t_{\rm Q}\sim 10^6$~yr causes significant tension with the inferred masses of SMBHs at these redshifts, since the presence of $\sim 10^{9}~M_\odot$ SMBHs at $z\sim6$ requires that quasars to shine continuously for nearly the entire Hubble time, i.e. $\sim 10^8-10^9~{\rm yr}$ \citep{Volonteri2010, Volonteri2012}. 
Hence, under the assumption of a light-bulb light curve we have to invoke very massive initial seeds ($M_{\rm seed}\sim 10^9\,M_{\sun}$), for which there is currently no
theoretical foundation nor any potential progenitor objects known, or
significantly super-Eddington accretion rates \citep{Du2015, Oogi2017}, which would however be in contrast to the low accretion rates inferred from X-ray observations, in order to explain the growth of the central billion solar mass black hole within the short quasar lifetime. 

However, there are several potential scenarios that could explain our findings. One option would be that the quasar has been growing for a much longer time than its lifetime estimate derived from the size of its proximity zone suggests, but it has been growing in a highly obscured phase, such that its ultraviolet (UV) radiation (and hence the ionizing continuum) has only ``broken out'' of this obscuring medium $\sim 10^4\,$yr ago \citep[e.g.][]{Hopkins2005, DiPompeo2017, Mitra2018}. In accordance with this scenario \citet{Sanders1988} proposed that ultraluminous infrared galaxies are the initial, heavily obscured stages of a quasar, which are revealed in the optical as an unobscured quasar at the end of the incipient dust-enshrouded phase. 
If this scenario is correct, and obscured quasars are an evolutionary rather than a geometric phenomenon, then the short quasar lifetime implied by $\rm J1335+3533$ implies a very large population of obscured quasars present in the high redshift universe: assuming a fiducial quasar lifetime of at least $t_{\rm Q, fiducial}\sim 10^8$~yr to grow the SMBH, the inferred UV luminous, i.e. unobscured, quasar lifetime of $t_{\rm Q}\sim 10^6$~yr from our proximity zone measurements, leads to an obscured quasar population that is at least $\sim 100$ times larger than the known quasar population, which would not have been discovered yet by any survey. However, this result would be in strong contrast to the estimated fraction of obscured quasars, which is much lower \citep{Polletta2008}. 

Instead of a light-bulb light curve model we could also assume a flickering light curve model in which the UV continuum emission of the quasar is allowed to vary, which has been suggested for instance in the context of changing-look quasars \citep[e.g.][]{LaMassa2015}. 
The IGM can probe quasar variability on timescales comparable to the equilibration timescale of the gas, i.e. $t_{\rm eq}\approx \Gamma_{\rm HI}^{-1}\approx 3\times 10^4\,\rm yr$, because of its finite response time to the quasar's radiation. The equilibration timescale can be inferred from the time evolution of the neutral gas fraction $x_{\rm HI}$ of the IGM, which evolves as
\begin{equation}
\frac{dx_{\rm HI}}{dt} = -\Gamma_{\rm HI}\,x_{\rm HI} + \alpha n_e \,(1-x_{\rm HI}), \label{eq:dxdt}
\end{equation}
where $\alpha$ denotes the recombination rate, and $n_e$ the number density of free electrons in the IGM \citep[e.g.][]{Khrykin2016}. Eqn.~\ref{eq:dxdt} is solved by
\begin{equation}
x_{\rm HI}(t) = x_{\rm HI,\,eq} + (x_{\rm HI,\,0} - x_{\rm HI,\,eq})\cdot e^{-t/t_{\rm eq}}, 
\end{equation}
where $x_{\rm HI,\,0}$ is the neutral gas fraction before the quasar turns on and $x_{\rm HI,\,eq} = \frac{\alpha n_e}{\Gamma_{\rm HI} + \alpha n_e}$ denotes the neutral gas fraction once the IGM is in ionizing equilibrium with the quasar's radiation. 

In a flickering quasar model the variable light curves could either result from intrinsic variations in the accretion rates, in which case active quasar phases would alternate with quiescent galaxy phases, or from variable obscuration by clouds in the BLR passing through our line-of-sight. 
Assuming an intrinsic variability of quasars due to changes in the accretion rate, our findings would imply episodic quasar lifetimes of $t_{\rm episodic}\sim 10^6~$yr to explain the occasional young quasars with small proximity zones, since the proximity zone measurement is only sensitive to the last active quasar phase, and any evidence of previously active phases in which the SMBH could have grown would be lost. However, a flickering quasar model will also reduce the average proximity zone size (Davies et al. in prep.), which is currently very well reproduced by a light-bulb model.

In order to distinguish between these different black hole formation scenarios we present an idea based on observations of nebular emission lines (such as H$\alpha$, H$\beta$, or [\ion{O}{3}]$\lambda 5007${\AA}) from ionized gas in the extended narrow line region (ENLR) with an integral field spectrograph. 
In the aforementioned scenarios, one scenario (scenario B, right panels of Fig.~\ref{fig:jwst}) predicts that the quasar's UV luminous phase just turned on $\sim 10^4$~yr ago, possibly after an initial, totally obscuring, dust-enshrouded phase, or -- under the assumption of a flickering quasar light curve -- it could have also turned on again after a quiescent phase. 
In the other scenario (scenario A, right panels of Fig.~\ref{fig:jwst}) the quasar has been UV obscured only along our line-of-sight due to clouds and variations in the BLR, but has indeed been shining in other directions and growing its SMBH for a much longer time. 

For either case we would expect to see a different surface brightness profile: If the quasar has indeed been UV obscured along our sight line, but emitting UV ionizing radiation into other directions for a very long time, i.e. $t_{\rm Q}\gtrsim 10^5\,$yr, we would expect to see an extended surface brightness profile. However, if the quasar's UV luminous phase turned on (again) only $t_{\rm Q}\sim 10^4\,$yr ago and a flickering quasar model, a dust-enshrouded initial stage, or possible super-Eddington accretion rates could explain the rapid growth of its SMBH, we expect to observe a surface brightness profile of its ENLR that is truncated at a distance of $R\sim c\times 10^4\rm\,yr\approx 3\,$ proper kpc (pkpc), which corresponds to the distance the quasar's radiation could have traveled during its lifetime. 

For calculating the two different surface brightness profiles we assumed the typical surface brightness of [\ion{O}{3}] around a quasar in the local universe from \citep{Liu2014} and re-scaled it by $(1+z)^{-4}$ to account for cosmological dimming. We choose the [\ion{O}{3}]$\lambda 5007${\AA} emission line doublet as a suitable tracer of the ionized ENLR, because it can be observed with NIRSpec on JWST at $z\sim6$ and traces relatively dense regions $n_{\rm H}\sim 10^3-10^5\,$cm$^{-3}$. Thus it will be almost instantaneously collisionally excited once the gas is photoionized to $\rm O^{2+}$. If the quasar shuts off the timescale for \ion{O}{3} to recombine is relatively short, i.e. $t_{\rm rec}\sim 10-100$~yr\footnote{The recombination time scale is $t_{\rm rec}\sim \frac{1}{\alpha\,n_e}$, where $\alpha=1.99\times 10^{-12}\,\rm cm^3\,s^{-1}$ is the recombination rate coefficient for \ion{O}{3} to recombine to \ion{0}{2} \citep{Draine}, and $n_e\sim 10^3-10^5\,$cm$^{-3}$ is the electron density. }
and hence even in the case of intermittent quasar activity the [\ion{O}{3}] emission from a previously active phase will have faded away if the quasar has been inactive for $t_{\rm off} > t_{\rm rec}$. Thus
the two aforementioned scenarios A and B will produce distinct surface brightness profiles.

This work presents a new step towards understanding the formation and growth of the first SMBHs in the universe. Our recent discovery of a young quasar population poses an interesting challenge for the current picture of SMBH growth and has put more stringent constraints on current quasar and galaxy evolution theories. Our proposed observations of the ENLR with the upcoming capabilities of JWST promise to shed further light onto the growth rate and formation of the SMBH in the center of the quasars and potentially provide a window to new physics (if super-Eddington accretion rates need to be invoked to explain the rapid growth), new insights into quasar-galaxy co-evolution (if indeed a flickering in the accretion rate of the quasar could explain our findings), or might indicate a large new population of obscured quasars in the high-redshift universe. 

Further progress could be made by identifying larger samples of young quasars (even at lower redshift), which might elucidate the connection between WLQs and quasar youth (Eilers et al. in prep.). The damping wing feature observed in quasars at very high redshift $z\gtrsim 7$, where the surrounding IGM is still significantly neutral, provides a measurement of the lifetime \textit{integrated} over all active phases and hence a statistical sample of high redshift quasars will provide further insights (Davies et al. in prep.), as well as observations of the proximity zones in the \ion{He}{2} Ly$\alpha$ forest, which has a longer equilibration timescale than hydrogen and is hence sensitive to longer quasar lifetimes (Khrykin et al. in prep.). Additionally a detailed modeling of proximity zones in the context of flickering light curve models is needed to interpret the observations (Davies et al. in prep.). 

\acknowledgments

The authors would like to thank Michele Fumagalli for sharing his composite spectrum of the LLSs, and Feige Wang, Bram Venemans, and Roberto Decarli for very helpful discussions and feedback. 

Some of the data presented in this paper were obtained at the W.M. Keck
Observatory, which is operated as a scientific partnership among the
California Institute of Technology, the University of California and
the National Aeronautics and Space Administration. The Observatory was
made possible by the generous financial support of the W.M. Keck
Foundation. 

The authors wish to recognize and acknowledge the very significant cultural role and reverence that the summit of Mauna Kea has always had within the indigenous Hawaiian community. We are most fortunate to have the opportunity to conduct observations from this mountain. 

This paper contains data based on observations obtained at the Gemini Observatory, acquired through the Gemini Observatory Archive, which is operated by the Association of Universities for Research in Astronomy, Inc., under a cooperative agreement with the NSF on behalf of the Gemini partnership: the National Science Foundation (United States), the National Research Council (Canada), CONICYT (Chile), Ministerio de Ciencia, Tecnolog\'{i}a e Innovaci\'{o}n Productiva (Argentina), and Minist\'{e}rio da Ci\^{e}ncia, Tecnologia e Inova\c{c}\~{a}o (Brazil). 

\bibliography{literatur_hz}


\end{document}